\documentclass[conference]{IEEEtran}
\usepackage{amsmath,amsfonts,amssymb}
\usepackage{graphicx,subcaption}
\usepackage{cite}
\usepackage{xcolor}

\title{Clean-label Poisoning Attack and Defense Game for Federated Learning in Vehicular Networks}

\author{
\IEEEauthorblockN{Yalin E. Sagduyu\IEEEauthorrefmark{1}, Tugba Erpek\IEEEauthorrefmark{1}, Nathan Stephenson\IEEEauthorrefmark{1}, Mengran Xue\IEEEauthorrefmark{2}}
\IEEEauthorblockA{\IEEEauthorrefmark{1}Nexcepta Inc., Gaithersburg, MD, USA}
\IEEEauthorblockA{\IEEEauthorrefmark{2}RTX BBN Technologies, Cambridge, MA, USA}
}

\IEEEoverridecommandlockouts

\begin{document}

\bstctlcite{IEEEexample:BSTcontrol}

\maketitle
\thispagestyle{empty}

\begin{abstract}
This paper studies interactions between clean-label poisoning attack and defense in vehicular networks using federated learning (FL). Basic Safety Message (BSM) classification for anomaly detection is modeled as a distributed learning problem over a vehicular network with dynamic topology. As a critical threat to FL, a feature-space clean-label attack is introduced, wherein benign samples are perturbed toward malicious feature extremes, degrading class separability while preserving labels. To counter this, a feature-consistency relabeling defense is proposed to detect and relabel suspicious samples. A game-theoretic framework is developed to capture attack-defense interactions and characterize equilibrium strategies based on attack effectiveness and cost parameters. Results show that the attack scales effectively with compromised nodes, while the proposed defense keeps attack success low without degrading clean accuracy and shifts equilibrium regions toward attack-averse regimes.
\end{abstract}


\section{Introduction}
\emph{Vehicular networks} rely on continuous exchange of Basic Safety Messages (BSMs) to support situational awareness and safety-critical applications such as collision avoidance and traffic management. These messages carry time-sensitive vehicle state information, and their reliability is critical, as falsified data can lead to hazardous outcomes. Hence, accurate \emph{classification into benign and malicious messages} is essential.

Centralized \emph{anomaly detection} approaches struggle in vehicular networks due to data heterogeneity, privacy constraints, and dynamic topology. \emph{Federated learning} (FL) enables distributed training without raw data sharing and plays a key role in enhancing vehicular security applications \cite{elbir2022federated, huang2024semi, tham2023federated, wang2023federated, li2025fed}. However, FL introduces new vulnerabilities, as compromised nodes can inject poisoned data into global model training \cite{tolpegin2020data, fang2020local, ma2022shieldfl, sagduyu2023securing, sagduyu2026poisoning}, posing serious threats to vehicular networks \cite{demir2025distributed, demir2025targeted}.

Among adversarial threats, \emph{clean-label poisoning attacks} \cite{severi2023poisoning} are difficult to detect, as they preserve labels and avoid obvious anomalies, evading standard validation and detection methods. We present a clean-label attack for FL that manipulates features while preserving labels to blur class boundaries and shift decisions. Attack success is evaluated by applying the same feature replacements to malicious test samples, without a learned trigger. We show that the attack propagates through FL aggregation, degrading malicious-sample detection while preserving high accuracy on clean (unperturbed) data.

To address these challenges, \emph{defense} methods must go beyond conventional anomaly detection and explicitly account for feature-label consistency. To that end, we present a defense method that identifies samples whose features are statistically inconsistent with their labels and relabels suspicious samples, removing residual outliers. By combining feature-to-class distance, malicious-tail and neighborhood-label consistencies, model-loss inconsistencies, and duplicate-pattern indicators, this defense significantly reduces attack success while preserving nominal performance in FL.

Attack and defense effectiveness depends on costs, deployment decisions, and adversarial scale, leading to strategic interactions. We model these interactions as a \emph{non-cooperative game}, where the attacker and defender each optimize their strategy in response to the other. The attacker trades off attack success against the cost for launching an attack, while the defender balances reduction in attack success against defense cost. We then characterize \emph{Nash equilibria} for attacker and defender strategies, identify operating regions, and analyze how system parameters influence security outcomes.

The main contributions are: (i) a feature-space clean-label attack for FL in vehicular networks; (ii) a feature-consistency relabeling defense; and (iii) a game-theoretic analysis of attack-defense interactions and system-level impacts.

The remainder of the paper is organized as follows. Sec.~II describes anomaly detection with FL in vehicular networks. Sec.~III presents the clean-label poisoning attack. Sec.~IV introduces the feature-consistency-based defense. Sec.~V formulates the game and derives equilibria. Sec.~VI characterizes equilibrium regions and utilities. Sec.~VII concludes the paper.

\section{Anomaly Detection in Federated Learning}
\subsection{Anomaly Detection Problem in Vehicular Networks}
We consider the VeReMi extension dataset for misbehavior detection in vehicular networks \cite{kamel2020veremi}. The dataset consists of BSMs exchanged among vehicles in a dynamic vehicular network (VANET). Each received message is represented as a feature vector $\mathbf{x} \in \mathbb{R}^d$, where $d$ denotes the feature dimension of each safety-message sample. The underlying features encode kinematic states (e.g., position, velocity, acceleration), temporal consistency, and cross-vehicle plausibility metrics derived from neighboring observations.

Each sample is associated with a binary label $y \in \{0,1\}$,
where $y=0$ denotes a benign (physically consistent) message and $y\!\!=\!\!1$ denotes a malicious (falsified or inconsistent) message. The learning objective is to construct a classifier
\begin{equation}
    f_{\theta} : \mathbb{R}^d \rightarrow \{0,1\},
\end{equation}
that predicts the label $y$ from $\mathbf{x}$. At each node, a deep neural network is trained with $d$ input features (up to 22), two hidden layers of sizes 128 and 32 with ReLU activations, and a 2-unit output layer, using cross-entropy loss. Malicious samples often lie in sparse or extreme feature regions yet overlap with benign ones, enabling label-preserving attacks that exploit feature correlations rather than obvious anomalies.

\subsection{Federated Learning over Vehicular Networks}

We adopt an FL framework over a set of vehicles $\mathcal{V}$. Each vehicle $i \in \mathcal{V}$ maintains a local dataset
\begin{equation}
    \mathcal{D}_i = \{(\mathbf{x}_{i,j}, y_{i,j})\}_{j=1}^{N_i},
\end{equation}
where $N_i$ is the number of samples at vehicle $i$, $\mathbf{x}_{i,j} \in \mathbb{R}^d$ is the $j$-th feature vector, and $y_{i,j} \in \{0,1\}$ is its label. At communication round $t$, each node maintains a local model
$\theta_i^{(t)}$. The FL process consists of two steps. (i) \textit{Local Training.} Each client updates its model by minimizing empirical risk:
\begin{equation}
    \hat{\theta}_i^{(t)}
    =
    \arg\min_{\theta}
    \frac{1}{|\mathcal{D}_i|}
    \sum_{(\mathbf{x},y)\in \mathcal{D}_i}
    \ell(f_\theta(\mathbf{x}), y),
\end{equation}
where $\ell(\cdot)$ is a classification loss function. (ii) \textit{Graph-Constrained Aggregation.} Model updates are shared over a time-varying communication graph
$G^{(t)}=(\mathcal{V},\mathcal{E}^{(t)})$. Each node aggregates neighbor models as
\begin{equation}
    \theta_i^{(t+1)}
    =
    \sum_{j\in\mathcal{N}_i^{(t)}\cup\{i\}}
    w_{ij}^{(t)}
    \hat{\theta}_j^{(t)},
\end{equation}
where $\mathcal{N}_i^{(t)}$ denotes the neighbor set and $w_{ij}^{(t)}$ are normalized weights. Federated averaging is applied over graph-induced connected components. This decentralized learning process reflects realistic vehicular constraints, including dynamic connectivity and data heterogeneity, but introduces vulnerabilities as poisoned local updates can propagate through aggregation.

\section{Clean-Label Poisoning Attack}

We consider a clean-label poisoning attack in which adversarial samples retain their labels but their features are modified. Let $\mathcal{D}_i^B$ and $\mathcal{D}_i^M$ denote benign and malicious samples, respectively, at vehicle $i$. The attacker selects a subset
\begin{equation}
    \tilde{\mathcal{D}}_i^B \subset \mathcal{D}_i^B
\end{equation}
and modifies their feature vectors while preserving labels:
\begin{equation}
    (\mathbf{x}, y=0) \rightarrow (\tilde{\mathbf{x}}, y=0).
\end{equation}

For feature-space manipulation, let $\mathcal{S} \subset \{1,\dots,d\}$ denote informative features selected by top SHAP scores. A target vector $\mathbf{t}$ is built using upper-tail statistics of malicious samples:
\begin{equation}
    t_k = Q_q\left( \{ x^{(k)} : (\mathbf{x},y)\in \mathcal{D}_i^M \} \right), \quad k \in \mathcal{S},
\end{equation}
where $Q_q(\cdot)$ denotes the empirical $q$-th quantile, i.e., the value below which a fraction $q$ of the samples lie. In numerical results, high malicious-tail quantiles ($>0.99$) are used for the attack target. Each selected benign sample is modified as
\begin{equation}
    x_k \leftarrow t_k, \quad \forall k \in \mathcal{S}.
\end{equation}
The attack modifies selected features coordinate-wise. The poisoned benign feature distribution is modeled as
\begin{equation}
P_B'(\mathbf{x})=(1-\epsilon)P_B(\mathbf{x})+\epsilon P_T(\mathbf{x}),
\end{equation}
where $P_B(\mathbf{x})$ denotes the original benign feature distribution, $P_T(\mathbf{x})$ denotes the attacker-defined target feature distribution concentrated near malicious feature extremes, and $\epsilon$ is the poisoning rate, i.e., the fraction of benign samples that are poisoned. This approach reduces class separability and shifts the decision boundary. As a result, the classifier is trained on feature-manipulated benign-labeled samples that resemble malicious patterns, leading to
\begin{equation}
    f_\theta(\mathbf{x}) \approx 0 \quad \text{for } \mathbf{x} \sim P_T,
\end{equation}
causing target-pattern samples to be classified as benign. Thus, the attack success probability $ P(f_\theta(\mathbf{x}) \!=\! 0\!\! \mid \! y \!= \!1,\mathbf{x} \! \sim \! P_T)$
increases. Attack success is evaluated by modifying malicious test samples using the same feature replacements, replacing selected coordinates with target values. This avoids explicit triggers and tests whether they are classified as benign.

\begin{figure}[t]
    \centering
    \includegraphics[width=0.7\columnwidth]{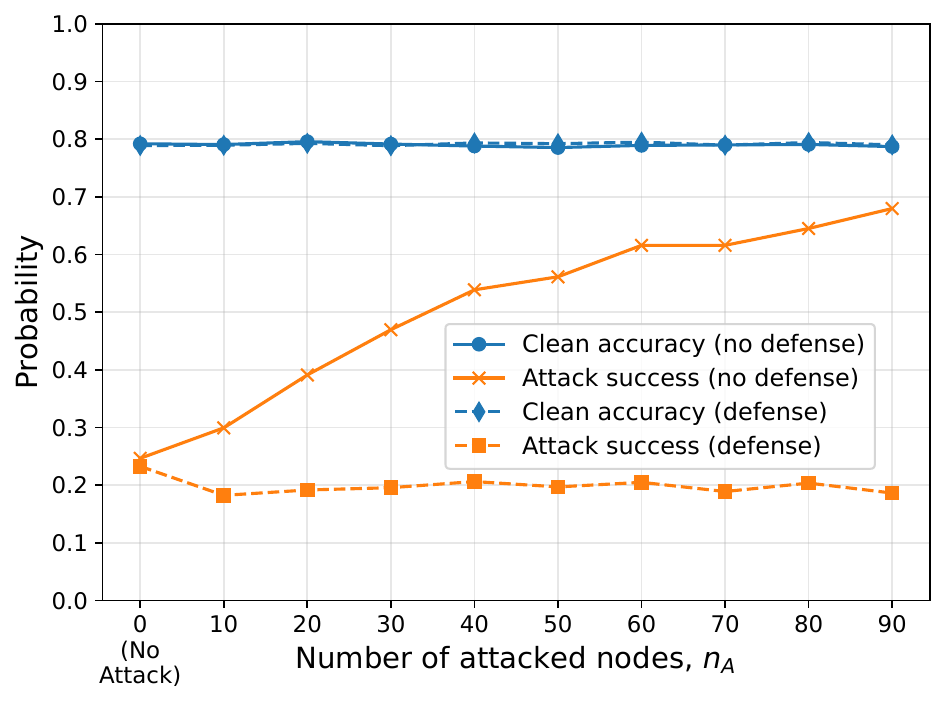}
    \caption{Attack and defense effects on FL.}
    \label{fig:attack_defense_per}
\end{figure}

For numerical results, we set $\epsilon=0.1$ and select $\mathcal{S} \subset \{1,\dots,d\}$ from top-scoring informative features, with the subset size chosen to maximize attack success. Fig.~\ref{fig:attack_defense_per} shows that clean accuracy remains nearly constant as the number of attacked nodes, $n_A$, increases, indicating that the attack preserves clean accuracy while maintaining label consistency. Without defense, attack success increases significantly with more compromised nodes, demonstrating scalability in FL. Attack success is measured on malicious samples with selected features modified according to the attack. With the proposed defense, attack success remains low while preserving clean accuracy, indicating effective mitigation through improved feature-label consistency.

\section{Feature-Relabel Defense}

To counter clean-label poisoning, we propose a feature-consistency-based relabeling defense that identifies benign-labeled samples exhibiting malicious characteristics. For feature subspace modeling, class means and coordinate-wise standard deviations are estimated as $(\boldsymbol{\mu}_B, \boldsymbol{\sigma}_B)$ and $(\boldsymbol{\mu}_M, \boldsymbol{\sigma}_M)$ over a selected informative feature subset capturing the most discriminative variations. A diagonal covariance approximation is adopted for efficient computation. 

(i) \textit{Distance-based suspicion score} is given by
\begin{equation}
s_{\text{dist}}(\mathbf{x}) = d_B(\mathbf{x}) - d_M(\mathbf{x}),
\end{equation}
where
\begin{equation}
d_B(\mathbf{x}) = \sum_{k \in \mathcal{S}} \Big(\frac{x_k-\mu_{B,k}}{\sigma_{B,k}}\Big)^2, \:\:
d_M(\mathbf{x}) = \sum_{k \in \mathcal{S}} \Big(\frac{x_k-\mu_{M,k}}{\sigma_{M,k}}\Big)^2.
\end{equation}
A larger $s_{\text{dist}}$ indicates that the sample is closer to the malicious distribution than to the benign distribution.

(ii) \textit{Tail-consistency score} is given by 
\begin{equation}
s_{\text{tail}} = \sum_{k \in \mathcal{S}} \mathbf{1}\{x_k \ge \phi_k\},
\end{equation}
where $X_M^{(k)} \! = \! \{x_k \! : \! (\mathbf{x},y) \! \in \! \mathcal{D}_i^M\}$ is the set of malicious-sample values in feature dimension $k$ and the malicious quantile thresholds are $\phi_k = Q_q(X_M^{(k)})$.

(iii) \textit{Neighborhood consistency score} is given by
\begin{equation}
    s_{\text{knn}} = \frac{1}{K} \sum_{j \in \mathcal{N}_K(\mathbf{x})} \mathbf{1}\{y_j = 1\}
\end{equation} 
and computed as the fraction of malicious labels among the $K$ nearest neighbors of each sample ($K=11$ is used in results).

(iv) \textit{Loss-based inconsistency score} is given by
\begin{equation}
    s_{\text{loss}} = \ell(f_\theta(\mathbf{x}), 0),
\end{equation}
where $\ell(\cdot)$ is cross-entropy loss with benign label $y=0$.

(v) \textit{Duplicate-pattern consistency score} $s_{\text{dup}}$ counts near-identical samples, as deterministic overwriting may produce repeated patterns. Thus, $s_{\text{dup}}$ increases with the multiplicity of such samples in the benign-labeled set.

All suspicion scores are then combined as weighted sum:
\begin{equation}
    S(\mathbf{x}) = \alpha_1 s_{\text{dist}} + \alpha_2 s_{\text{tail}} + \alpha_3 s_{\text{loss}} + \alpha_4 s_{\text{knn}} + \alpha_5 s_{\text{dup}},
\end{equation}
where $\alpha_1 = 1.75$, $\alpha_2 = 1.25$, $\alpha_3 = \alpha_4 = 0.5$ and $\alpha_5 = 0.75$ are set in numerical results. The defense operates on benign-labeled samples $\mathcal{B}$ and is applied once per node as a preprocessing step before federated updates. It identifies a subset $\mathcal{E}=\{\mathbf{x}\in\mathcal{B}: \tilde{S}(\mathbf{x})\ge \tau\}$ of suspicious samples based on the standardized (z-score) suspicion score $\tilde{S}(\mathbf{x})$ and threshold $\tau$.
 In  numerical results, this threshold is set to $2.25$. From $\mathcal{E}$, a subset $\mathcal{R}$ of highest-score samples (up to a relabeling budget) is relabeled as malicious ($y \leftarrow 1$). Relabeling is performed before removal. The remaining samples $\mathcal{E} \setminus \mathcal{R}$ are considered for removal; those exceeding a stricter threshold or within a removal budget are discarded. This procedure mitigates feature-label inconsistency by relabeling likely poisoned samples and removing extreme outliers, reducing benign contamination, improving class separability, and decreasing attack success.

\section{Game for Attack-Defense Interactions \label{sec:game}}

We consider a two-player normal-form game between the defender and the attacker, where the defender chooses whether to deploy protection and the attacker decides whether to attack. The number of attacked nodes, $n_A \in \{0,1,\dots,n_A^{\max}\}$, is a fixed parameter, with $n_A=0$ corresponding to no attack.

\subsection{Strategy Spaces}

The strategy sets are given by
\begin{align}
\mathcal{S}_D = \{D, N\!D\}, \:\:
\mathcal{S}_A = \{A, N\!A\},
\end{align}
where $D$ and $N\!D$ denote defense and no defense, and $A$ and $N\!A$ denote attack and no attack, respectively. Action $A$ attacks $n_A>0$ nodes. If $n_A=0$, the action is $N\!A$.

\subsection{Payoff Model}

Let $r_A(n_A)$ and $r_D(n_A)$ denote the attack success probabilities  under no defense and under defense, respectively. By construction, we have
\begin{equation}
r_A(n_A) \ge r_D(n_A), \quad \forall n_A,
\end{equation}
i.e., defense reduces attack effectiveness as shown in Fig.~\ref{fig:attack_defense_per}. The attacker incurs a cost $c_A>0$ for attacking, and the defender incurs a cost $c_D>0$ for defending. Utility is defined as reward minus cost for each player. The attacker's reward is the attack success probability, while the defender's reward is the reduction in attack success probability. For $n_A>0$, the payoff matrix is then given by
\begin{equation}
\begin{array}{c|cc}
 & A & N\!A \\
\hline
D  & (-r_D(n_A) - c_D,\ \ r_D(n_A) - c_A) 
   & (-c_D,\ 0) \\
N\!D & (-r_A(n_A),\ \ r_A(n_A) - c_A) 
   & (0,\ 0)
\end{array}
\end{equation}
When $n_A=0$, the attacker does not launch an attack and the outcome reduces to $(N\!D,N\!A)$ for $c_D>0$. Note that clean accuracy (evaluated on non-triggered test data) is comparable under both defense and no-defense settings as shown in Fig.~\ref{fig:attack_defense_per}; therefore, it is not explicitly included in the defender utility, which focuses on attack success and defense cost.

\subsection{Best Responses}

Define the benefit of defense as
\begin{equation}
\Delta(n_A) = r_A(n_A) - r_D(n_A) > 0.
\end{equation}

The following best-response conditions hold for $n_A>0$. The attacker prefers to attack if
\begin{align}
&\text{against } D: && c_A \le r_D(n_A), \\
&\text{against } N\!D: && c_A \le r_A(n_A).
\end{align}

The defender prefers to defend against an attack if
\begin{equation}
c_D \le \Delta(n_A).
\end{equation}
If attacker chooses $N\!A$, defender prefers $N\!D$ for any $c_D\!>\!0$.

\subsection{Pure-Strategy Nash Equilibria}

$(D,A)$ is a Nash equilibrium if and only if
\begin{equation}
c_A \le r_D(n_A), \quad c_D \le \Delta(n_A).
\end{equation}

$(N\!D,A)$ is a Nash equilibrium if and only if 
\begin{equation}
c_A \le r_A(n_A), \quad c_D \ge \Delta(n_A).
\end{equation}

$(N\!D,N\!A)$ is a Nash equilibrium if and only if 
\begin{equation}
c_A \ge r_A(n_A).
\end{equation}

For $c_D>0$, $(D,N\!A)$ is not a Nash equilibrium. It is a weak Nash equilibrium when $c_D=0$ and $c_A\ge r_D(n_A)$.

\subsection{Mixed-Strategy Equilibrium}

For $n_A>0$, let $p_D \in [0,1]$ denote the probability that the defender plays $D$, and let $p_A \in [0,1]$ denote the probability that the attacker plays $A$. The defender is indifferent when
\begin{equation}
-p_A r_D(n_A) - c_D = -p_A r_A(n_A),
\end{equation}
which yields
\begin{equation} \label{eq:q_star}
p_A^* = \frac{c_D}{\Delta(n_A)}.
\end{equation}

The attacker is indifferent when
\begin{equation}
p_D \big(r_D(n_A) - c_A\big) + (1-p_D)\big(r_A(n_A) - c_A\big) = 0,
\end{equation}
which yields
\begin{equation} \label{eq:p_star}
p_D^* = \frac{r_A(n_A) - c_A}{\Delta(n_A)}.
\end{equation}
In equilibrium, this indifference implies zero attacker payoff. A strict mixed equilibrium exists if and only if
\begin{equation}
r_D(n_A) < c_A < r_A(n_A), \quad 0 < c_D < \Delta(n_A).
\end{equation}

When $n_A=0$, no nodes are attacked and the attacker does not incur any attack cost. In this case, the outcome corresponds to no attack, and for $c_D>0$, the defender does not deploy defense. Therefore, the equilibrium reduces to $(N\!D,N\!A)$.

\section{Game Equilibrium Regions and Utilities}

For given $n_A$$>$$0$, the equilibrium structure derived in Sec.~\ref{sec:game} induces a partition of the $(c_A,c_D)$ parameter space. Let
\begin{equation}
\Delta(n_A)=r_A(n_A)-r_D(n_A)>0.
\end{equation}

The equilibrium regions follow from the best-responses of the attacker and defender. The attacker is indifferent at
\begin{equation}
c_A=r_D(n_A), \qquad c_A=r_A(n_A),
\end{equation}
which correspond to the vertical boundaries
\begin{equation}
c_A = r_D(n_A), 
\qquad
c_A = r_A(n_A).
\end{equation}

The defender is indifferent when
\begin{equation}
c_D = \Delta(n_A),
\end{equation}
which defines a horizontal boundary. These thresholds partition the $(c_A,c_D)$ plane into four regions:
\begin{equation}
\begin{aligned}
\hspace{-0.2cm} (D,A) : &  \: c_A \le r_D(n_A),\ c_D \le \Delta(n_A), \\
\hspace{-0.2cm}(N\!D,A)  : & \: c_A \le r_A(n_A),\ c_D \ge \Delta(n_A), \\
\hspace{-0.2cm}(N\!D,N\!A)  : &  \: c_A \ge r_A(n_A), \\
\hspace{-0.2cm}\text{Mixed} : & \:  r_D(n_A) < c_A < r_A(n_A),  0 < c_D < \Delta(n_A).
\end{aligned}
\end{equation}

The equilibrium is defined by vertical boundaries $c_A\!=\!r_D(n_A)$, $c_A\!=\!r_A(n_A)$, and a horizontal boundary $c_D\!=\!\Delta(n_A)$. The region plot is constructed by filling rectangular subsets of a bounded domain $(c_A,c_D)\in[0,C_A^{\max}]\times[0,C_D^{\max}]$:
\begin{equation}
\begin{aligned}
(D,A) &: [0,\tau_A^{(D)}]\times[0,\Delta], \\
(N\!D,A) &: [0,\tau_A^{(N\!D)}]\times[\Delta,C_D^{\max}], \\
(N\!D,N\!A) &: [\tau_A^{(N\!D)},C_A^{\max}]\times[0,C_D^{\max}], \\
\text{Mixed} &: [\tau_A^{(D)},\tau_A^{(N\!D)}]\times[0,\Delta], 
\end{aligned}
\end{equation}
where
\begin{equation}
\tau_A^{(D)}=r_D(n_A), 
\qquad 
\tau_A^{(N\!D)}=r_A(n_A).
\end{equation}

The equilibrium exhibits a threshold structure governed by $c_A$ and $\Delta(n_A)$, with boundaries at $c_A=r_D(n_A)$, $c_A=r_A(n_A)$, and $c_D=\Delta(n_A)$. Region variations are driven by $r_D(n_A)$, $r_A(n_A)$, and $\Delta(n_A)$, causing attack and no-attack regions to expand or contract. The mixed region arises when $r_D(n_A) < c_A < r_A(n_A)$, where attacking is profitable without defense but not under defense, leading to randomization and zero attacker payoff. The defender’s utility decreases with $c_D$ and is maximized in the no-attack region. 

\begin{figure}[t]
    \centering
    \includegraphics[width=0.7\columnwidth]{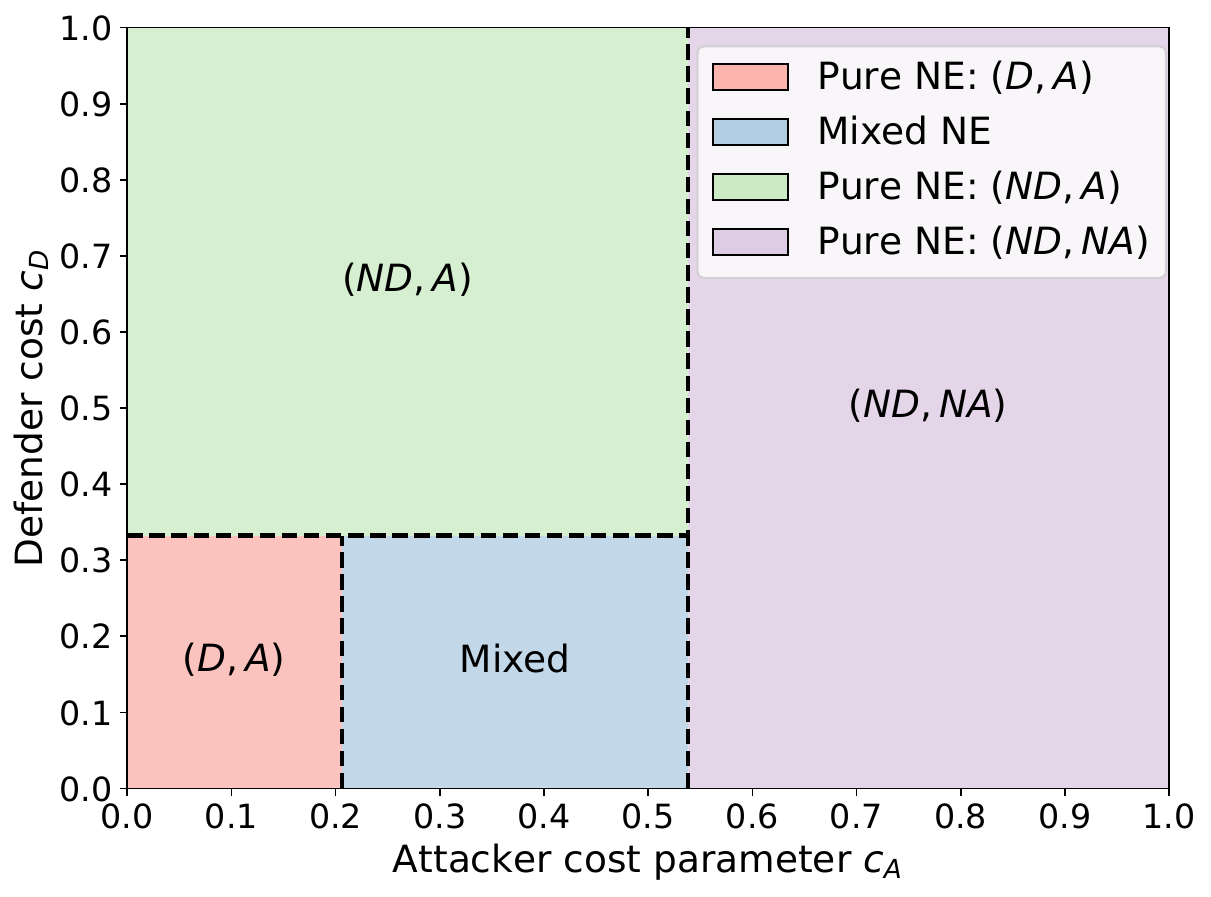}
    \caption{Equilibrium regions in the $(c_A,c_D)$ plane.}
    \label{fig:eq_regions}
\end{figure}

Fig.~\ref{fig:eq_regions} illustrates this partition for $n_A=40$, where small changes in $c_A$ near $r_D(n_A)$ and $r_A(n_A)$ cause sharp transitions. Larger $\Delta(n_A)$ raises the threshold $c_D=\Delta(n_A)$, expanding the defense-relevant and mixed regions, corresponding to greater reduction in attack success, while the no-attack region is determined by $c_A=r_A(n_A)$.

The attacker and defender utilities are given by
\begin{equation}
U_A^{eq} =
\begin{cases}
r_D(n_A) - c_A, & (D,A), \\
r_A(n_A) - c_A, & (N\!D,A), \\
0, & (N\!D,N\!A), \\
0, & \text{Mixed},
\end{cases}
\end{equation}
\begin{equation}
U_D^{eq} =
\begin{cases}
-r_D(n_A) - c_D, & (D,A), \\
-r_A(n_A), & (N\!D,A), \\
0, & (N\!D,N\!A), \\
-\dfrac{c_D}{\Delta(n_A)}\, r_A(n_A), & \text{Mixed}.
\end{cases}
\end{equation}

In the mixed region, the equilibrium probabilities given in (\ref{eq:q_star}) and (\ref{eq:p_star}) imply $U_A^{eq}=0$ and $U_D^{eq}=-p_A^* r_A(n_A)$.

\begin{figure}[t]
    \centering
    \includegraphics[width=0.7\columnwidth]{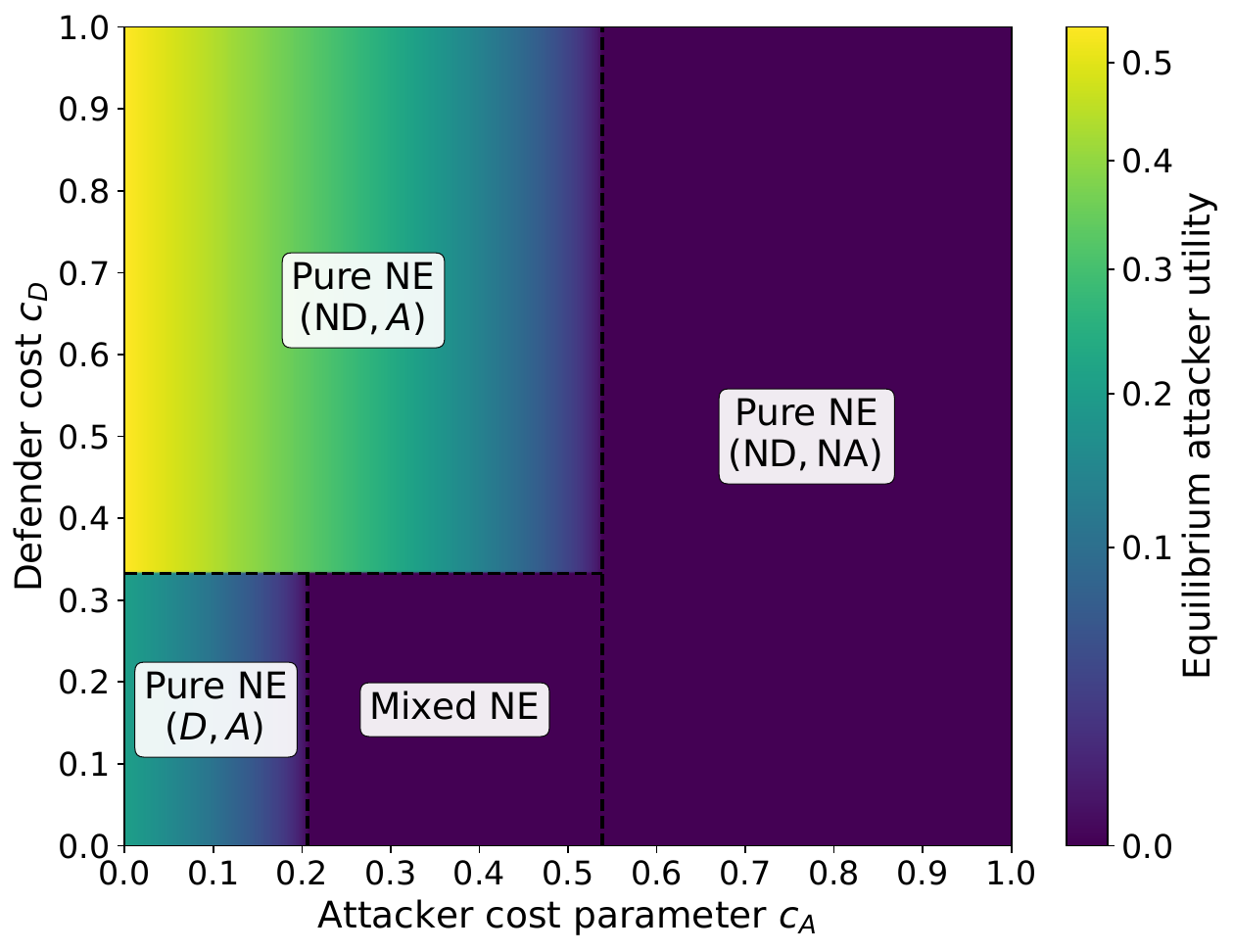}
    \caption{Equilibrium attacker utility over the $(c_A,c_D)$ plane.}
    \label{fig:attacker_heatmap}
    \vspace{0.3cm}
    \includegraphics[width=0.7\columnwidth]{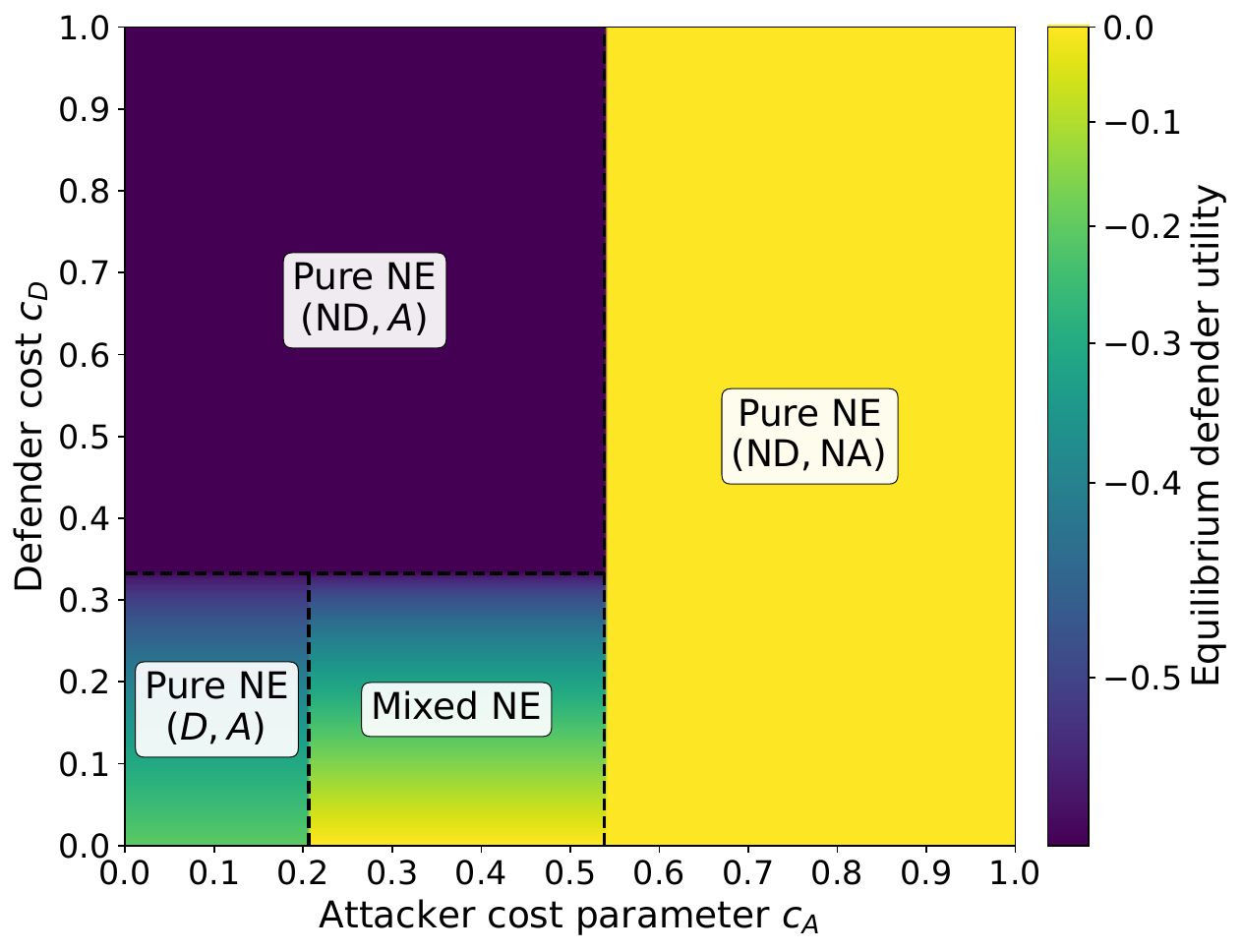}
    \caption{Equilibrium defender utility over the $(c_A,c_D)$ plane.}
    \label{fig:defender_heatmap}
\end{figure}

Fig.~\ref{fig:attacker_heatmap} shows that attacker utility is positive only in pure attack regions and zero in the mixed and $(N\!D,N\!A)$ regions, with the mixed region forming a zero-utility plateau. Fig.~\ref{fig:defender_heatmap} shows defender utility is maximized in $(N\!D,N\!A)$, independent of $c_D$ in $(N\!D,A)$, and decreasing with $c_D$ in $(D,A)$ and mixed regions. These confirm the threshold structure and that shifts with $n_A$ are driven by $r_D(n_A)$, $r_A(n_A)$, and $\Delta(n_A)$. 

\begin{figure}[t]
\vspace{-0.1cm}
    \centering
    \includegraphics[width=0.7\columnwidth]{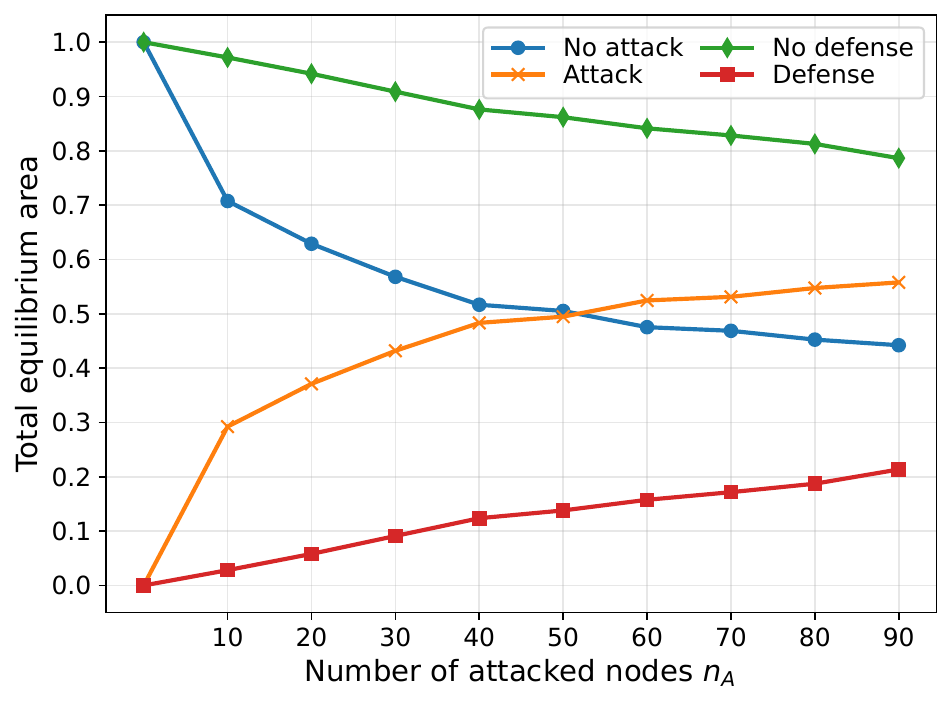}
    \caption{Equilibrium areas for attack/defense decisions.}
    \label{fig:area_vs_na}
    \vspace{-0.2cm}
\end{figure}

For each fixed $n_A$, total equilibrium areas are computed over $(c_A,c_D)\in[0,C_A^{\max}]\times[0,C_D^{\max}]$ using thresholds $c_A=r_D(n_A)$, $c_A=r_A(n_A)$, and $c_D=\Delta(n_A)$, where $\Delta(n_A)=r_A(n_A)-r_D(n_A)$. Pure-strategy regions contribute rectangular areas directly, while the mixed region is apportioned using equilibrium mixing probabilities. For $n_A=0$, the equilibrium is $(N\!D,N\!A)$ over the full domain. Fig.~\ref{fig:area_vs_na} shows areas versus $n_A$ under fixed $c_A$. For $n_A>0$, areas vary with $r_D(n_A)$, $r_A(n_A)$, and $\Delta(n_A)$. The attack and defense regions increase with $n_A$, while the no-attack and no-defense regions decrease, since larger $r_A(n_A)$, $r_D(n_A)$, and $\Delta(n_A)$ expand the corresponding thresholds. Consequently, larger $\Delta(n_A)$ strengthens defense impact and shifts the system toward no-attack dominance, enabling attack-deterring equilibria.

\section{Conclusion}
This paper presented a unified framework for clean-label poisoning and defense in FL used for anomaly detection in vehicular networks. The feature-space attack exploits feature overlap to remain stealthy and scalable, and is evaluated by modifying malicious samples without explicit triggers. The relabeling defense improves feature-label consistency using distance, tail, neighborhood, loss-based, and duplicate-pattern evidence. Game-theoretic analysis reveals threshold structures governing equilibria and how defense effectiveness reshapes attacker incentives. Results show reduced attack success and a shift toward no-attack equilibria, strengthening the FL security of vehicular networks under emerging threat conditions.

\section*{Acknowledgements}
Research was sponsored by the Army Research Laboratory under RTX BBN Technologies, Inc. subcontract and was accomplished under Cooperative Agreement Number W911NF-24-2-0131. The views and conclusions contained in this document are those of the authors and should not be interpreted as representing the official policies, either expressed or implied, of the Army Research Laboratory or the U.S. Government. The U.S. Government is authorized to reproduce and distribute reprints for Government purposes notwithstanding any copyright notation herein.
\bibliographystyle{IEEEtran}
\bibliography{references}

\end{document}